\documentclass[aps,prd,preprint,preprintnumbers,nofootinbib,nobibnotes,]{revtex4-1}

\pdfoutput=1

\usepackage{amssymb,amsmath}
\usepackage{color}
\usepackage{graphicx,subfigure}
\usepackage{hyperref}
\usepackage{float} 
\usepackage[T1]{fontenc}

\newcommand{\nn}{\nonumber}

\newcommand \ket[1]{
        \left| #1 \right>
}

\newcommand \bra[1]{
        \left< #1 \right|
}

\begin{document}

\title{Relations between $b\rightarrow c\tau \nu$ Decay Modes in Scalar Models}

\author{Stefan Schacht}
\email{ss3843@cornell.edu}
\affiliation{Department of Physics, LEPP, Cornell University, Ithaca, NY 14853, USA}
\author{Amarjit Soni}
\email{adlersoni@gmail.com}
\affiliation{Physics Department, Brookhaven National Laboratory, Upton, NY 11973, USA}

\begin{abstract}
As a consequence of the Ward identity for hadronic matrix elements, we find relations between the differential decay rates of semileptonic decay modes with the underlying quark-level transition $b\rightarrow c\tau \nu$, which are valid in scalar models. The decay-mode dependent scalar form factor is the only necessary theoretical ingredient for the relations. Otherwise, they combine measurable decay rates as a function of the invariant mass-squared of the lepton pair $q^2$ in such a way that a universal decay-mode independent function is found for decays to vector and pseudoscalar mesons, respectively. This can be applied to the decays $B\rightarrow D^{*}\tau\nu$, $B_s\rightarrow D_s^*\tau\nu$, $B_c\rightarrow J/\psi\tau\nu$ and $B\rightarrow D\tau\nu$, $B_s\rightarrow D_s\tau\nu$, $B_c\rightarrow \eta_c\tau\nu$, with implications for $R(D^{(*)})$, $R(D_s^{(*)})$, $R(J/\psi)$, $R(\eta_c)$, and $\mathcal{B}(B_c\rightarrow \tau\nu)$. The slope and curvature of the characteristic $q^2$-dependence is proportional to scalar new physics parameters, facilitating their straight forward extraction, complementary to global fits. 
\end{abstract}

\maketitle

\section{Introduction \label{sec:intro} }

There are by now several long-term tensions in flavor physics observables of underlying $b\rightarrow c\tau\nu$ transitions 
that hint for a violation of lepton-flavor universality (LFU) between light leptons $l=e,\mu$ and heavy $\tau$ leptons. 
Current experimental determinations of the ratios  
\begin{align}
R(\{V,P\}) \equiv \frac{\mathcal{B}(B_q\rightarrow \{V,P\} \tau\nu)}{\mathcal{B}(B_q\rightarrow \{V,P\} l\nu )}\,, 
\end{align} 
are provided by the Heavy Flavor Averaging Group (HFLAV)~\cite{Amhis:2019ckw, Lees:2012xj, Lees:2013uzd, Huschle:2015rga, Aaij:2015yra, Hirose:2016wfn, Hirose:2017dxl, Aaij:2017uff, Aaij:2017deq, Abdesselam:2019dgh},
\begin{align}
R(D^*)   &= 0.295 \pm 0.011 \pm 0.008\,, \label{eq:current-data-1} \\
R(D) &= 0.340 \pm 0.027 \pm 0.013\,, \label{eq:current-data-2} 
\end{align}
and are in tension with corresponding averages of SM predictions quoted by HFLAV as~\cite{Amhis:2019ckw, Bigi:2016mdz, Bernlochner:2017jka, Bigi:2017jbd, Jaiswal:2017rve}
\begin{align}
R(D^*)^{\mathrm{SM}} &= 0.258 \pm 0.005\,, \\
R(D)^{\mathrm{SM}}   &= 0.299 \pm 0.003\,.  
\end{align}
An updated SM prediction using additional data on decays to light leptons~\cite{Waheed:2018djm} is provided in Ref.~\cite{Gambino:2019sif}  
\begin{align}
R(D^*)^{\mathrm{SM}} = 0.254^{+0.007}_{-0.006}\,, \label{eq:SM-value}
\end{align}
see also Refs.~\cite{Jaiswal:2020wer, Iguro:2020cpg}.
There are further hadronic decays with the same underlying quark level transition like
$B_s\rightarrow D_s^{(*)}\tau\nu$, $B_c\rightarrow J/\psi\tau\nu$, $B_c\rightarrow \eta_c\tau \nu$,
as well as baryonic decays~\cite{Bernlochner:2018kxh,Bernlochner:2018bfn,Datta:2017aue, Detmold:2015aaa,Mannel:2015osa, Boer:2018vpx, Boer:2019zmp, Colangelo:2020vhu}. 
A $1.8\sigma$-tension has been seen in $B_c\rightarrow J/\psi \tau \nu$ decays
\begin{align}
R(J/\psi) &= 0.71\pm 0.17\pm 0.18\,, & \text{\cite{Aaij:2017tyk}} \label{eq:current-data-3} \\
R(J/\psi)^{\mathrm{SM}} &= 0.25\pm 0.03\,, &  \text{\cite{Cohen:2019zev}} \label{eq:RJpsi-SM} 
\end{align}
see also Refs.~\cite{Cohen:2018dgz, Murphy:2018sqg, Watanabe:2017mip, Issadykov:2018myx, Dutta:2017xmj, Wang:2018duy, Azizi:2019aaf, Leljak:2019eyw}.  
Analogous deviations are also seen in $b\rightarrow sl^+l^-$ decays, but there between muon and electron final states~\cite{Aaij:2014ora, Aaij:2017vbb, Aaij:2019wad, Abdesselam:2019wac, Hiller:2003js,  Huber:2005ig, Bobeth:2007dw, Bordone:2016gaq, Hiller:2014yaa, Hiller:2014ula,Aloni:2017ixa}, and there are interesting cross-correlations to high-$p_T$ physics~\cite{Afik:2020cvr,Borschensky:2020hot, Altmannshofer:2017poe, Iguro:2018fni}.
On top of these tensions with LFU, extractions of $V_{cb}$ and $V_{ub}$ from semileptonic decays differ when performed with inclusive and exclusive decays---a long-term story which we anticipate to continue to evolve in unexpected ways also in the 
future~\cite{Gambino:2019sif, Gambino:2016jkc, Bigi:2016mdz, Glattauer:2015teq, Aubert:2009ac, Caprini:1997mu, Abdesselam:2017kjf, Boyd:1997kz, Bigi:2017njr, Grinstein:2017nlq, Jaiswal:2017rve, Bernlochner:2017jka, Bigi:2017jbd, Dey:2019bgc, Waheed:2018djm, Bernlochner:2019ldg, Bernlochner:2020tfi, Bernlochner:2018bfn, Bernlochner:2018kxh, Bernlochner:2017xyx, Aaij:2020hsi, Aaij:2020xjy, Colangelo:2016ymy}. 
A lot of experimental improvement regarding semileptonic decays is expected in the future~\cite{Cerri:2018ypt, Kou:2018nap, Gambino:2020jvv}.
For progress, form factor results from  
lattice QCD~\cite{Bailey:2014tva, Lattice:2015rga, Na:2015kha,  Harrison:2017fmw, McLean:2019sds, Aviles-Casco:2019vin, Kaneko:2018mcr, Gambino:2020crt, Flynn:2019jbg, Flynn:2019any, Flynn:2016vej,Murphy:2018sqg,Colquhoun:2016osw, Heitger:2003nj, DellaMorte:2013ega,  DellaMorte:2015yda} 
and also LCSRs~\cite{Gubernari:2018wyi, Faller:2008tr, Bordone:2019vic, Bordone:2019guc} are very important. 

A lot of progress has been made in the research of the ability of new physics (NP) models, including from the beginning scalar models, to explain the data~\cite{Kamenik:2008tj, Fajfer:2012vx, Celis:2012dk, Tanaka:2012nw, Blanke:2018yud, Blanke:2019qrx, Freytsis:2015qca, Bhattacharya:2016zcw, Ivanov:2017mrj,  Alok:2017qsi, Bifani:2018zmi, Shi:2019gxi, Greljo:2015mma, Boucenna:2016wpr,  Boucenna:2016qad, Megias:2017ove, Li:2016vvp, Fajfer:2012jt, Deshpande:2012rr, Sakaki:2013bfa, Duraisamy:2014sna, Calibbi:2015kma, Fajfer:2015ycq, Barbieri:2015yvd, Alonso:2015sja, Bauer:2015knc, Das:2016vkr, Deshpand:2016cpw, Sahoo:2016pet, Dumont:2016xpj, Becirevic:2016yqi, Barbieri:2016las,  DiLuzio:2017vat, Chen:2017hir, Bordone:2017bld, Altmannshofer:2017poe, Becirevic:2018afm, Crivellin:2012ye, Celis:2012dk, Crivellin:2015hha, Celis:2016azn, Chen:2017eby, Iguro:2017ysu, Chen:2018hqy, Li:2018rax, Afik:2020cvr,  Altmannshofer:2020axr, Bar-Shalom:2018ure, Nandi:2016wlp,  Asadi:2018sym, Asadi:2018wea, Asadi:2019xrc, Alonso:2017ktd, Alonso:2016oyd, Alonso:2016gym, Bhattacharya:2020lfm, Bhattacharya:2019olg,Bhattacharya:2016mcc, Bhattacharya:2014wla, Bernlochner:2017jka, Ligeti:2016npd, Becirevic:2016hea, Becirevic:2019tpx, Biancofiore:2013ki, Colangelo:2018cnj, Martinez:2018ynq, Aloni:2018ipm,Aloni:2017eny, Deschamps:2009rh, Alok:2019uqc, Crivellin:2017zlb, Crivellin:2019dwb, Jaiswal:2020wer, Bhattacharya:2018kig, Iguro:2020cpg, Iguro:2018vqb, Iguro:2018qzf, Leljak:2019eyw, Bardhan:2016uhr, Azatov:2018knx, Bardhan:2019ljo, Bigaran:2019bqv, Gargalionis:2019drk, Cai:2017wry}.
An important way to probe for NP are relations between different decay modes. 
In non-leptonic decays this is a tool which is known for a long time, and there based on SU(3)$_F$ methods, 
see for example Refs.~\cite{Gronau:1990ka, Gronau:1994rj, Gronau:1995hm, Dery:2020lbc, Grossman:2018ptn, Hiller:2012xm, Muller:2015rna,Grossman:2012ry}.

For semileptonic $b\rightarrow c\tau\nu$ decays, model-specific relations that connect different decay modes are known for left-handed vector models as the relation~\cite{Bhattacharya:2014wla, Greljo:2015mma, Calibbi:2015kma, Boucenna:2016wpr, Boucenna:2016qad, Celis:2016azn} 
\begin{align}
\text{left-handed vector models:}\qquad \frac{R(V)}{R(V)^{\mathrm{SM}}} = \frac{R(P)}{R(P)^{\mathrm{SM}}} = \mathrm{const}. \quad \forall\,\, V, P\,,
\end{align} 
which is e.g.~also found in the $R$-parity violating SUSY model 
considered in Refs.~\cite{Altmannshofer:2017poe,Altmannshofer:2020axr}. 
No matter which decay channel is considered on the left-hand side, the same expression is obtained on the right-hand side.
In this paper, we present similar relations between differential decay rates of different decay modes in scalar models.
They can be found in Eqs.~(\ref{eq:sum-rule-10})--(\ref{eq:sum-rule-14}) and Fig.~\ref{fig:sumrules}.
The resulting decay-mode independent functions of the invariant lepton mass-squared $q^2$ are a finger print of the model: 
Its slope and curvature are directly proportional to NP parameters which can thus be readily extracted.
A departure from that characteristic function would be a sign of NP beyond scalar models. 

In contrast to non-leptonic sum rules, which are based on the approximate flavor symmetry of QCD, the relations that we consider 
here are based on ones between hadronic form factors which follow from the Ward identity, and are therefore exact. 
We do not use flavor symmetries to derive these relations.

Note that it is known to be 
challenging~\cite{Lees:2012xj,Fajfer:2012jt, Crivellin:2012ye, Crivellin:2013wna, Haller:2018nnx, Kou:2018nap} to explain the available $b\rightarrow c\tau\nu$ data with Two-Higgs-Doublet Models (2HDM)~\cite{Gunion:1989we, Kalinowski:1990ba, Hou:1992sy, Gunion:2002zf, Branco:2011iw, Haber:2013mia,Craig:2013hca,Gori:2017qwg, Grzadkowski:2018ohf, Celis:2012dk, Jung:2010ik, Pich:2009sp,Bernon:2015qea} of type I and II while respecting other constraints \cite{Akeroyd:2016ymd, Misiak:2017bgg,Aaboud:2018gjj, Haller:2018nnx, Misiak:2020ryv, Misiak:2020vlo, PDG2020,Enomoto:2015wbn,Cheng:2014ova,Arbey:2017gmh}.
Quark flavor constraints from $\mathcal{B}(B\rightarrow X_s\gamma)$, $\mathcal{B}(B_s^0\rightarrow \mu^+\mu^-)$, $\mathcal{B}(B\rightarrow \tau\nu)$ and others, without semileptonic $b\rightarrow c\tau\nu$ and $b\rightarrow sll$ decay modes, imply roughly~\cite{Haller:2018nnx} 
\begin{align}
\text{2HDM-I:}\quad & \text{$\tan\beta \gtrsim 2$ for $m_{H^{\pm}} \sim 300$ GeV and $\tan\beta \gtrsim 1.5$ for $m_{H^{\pm}} \sim 600$ GeV}\,, \nn\\
\text{2HDM-II:}\quad & \text{$m_{H^{\pm}} \gtrsim 600$ GeV and $\tan\beta \lesssim 25$ for $m_{H^{\pm}} \leq 1$ TeV\,,}\nn
\end{align}
see Ref.~\cite{Haller:2018nnx} for details. 
Especially important herein is the bound from $\mathcal{B}(B\rightarrow X_s\gamma)$~\cite{Misiak:2017bgg, Misiak:2020vlo}.
On the other hand, translating the allowed region of the model independent two-dimensional scalar global fit to $b\rightarrow c\tau\nu$ observables performed in Refs.~\cite{Blanke:2018yud, Blanke:2019qrx} into allowed $\tan\beta$ and $m_{H^{\pm}}$ values in the 2HDM-II, we obtain very small values of order $\tan\beta\lesssim 1$ and $m_{H^{\pm}}\lesssim 2\,\mathrm{GeV}$. 
That means the current measurements of $b\rightarrow c\tau \nu$ observables can only be explained simultaneously for parameter values clearly excluded by other bounds, e.g. $\mathcal{B}(B\rightarrow X_s\gamma)$.
This observation agrees with Fig.~6 in Ref.~\cite{Enomoto:2015wbn} where the allowed parameter space for explaining $R(D^*)$ and $R(D)$ also converges only for very small $\tan\beta$ and $m_{H^{\pm}}$, excluded by other data.
Applying the bounds from Ref.~\cite{Haller:2018nnx} to the 2HDM-I, the resulting Wilson coefficients~\cite{Celis:2012dk} 
are of the order $\vert C_R\vert \equiv \cot^2\beta m_b m_{\tau}/m^2_{H^{\pm}} \lesssim 2\cdot 10^{-5}$, far too small in order to account for either one of $R(D^{(*)})$~\cite{Enomoto:2015wbn}.
However, examples of more general 2HDMs with flavor-alignment exist that indeed can explain $R(D^{(*)})$~\cite{Celis:2012dk,Pich:2009sp,Jung:2010ik}.

For $b\rightarrow sll$ LFU ratios $R(K^{(*)})$, the Wilson coefficients $C_{9,10}$ play an important role, see for recent fits Ref.~\cite{Aebischer:2019mlg}. However, in the 2HDM-I or II the contributions to $C_{9,10}$ are suppressed by $\cot^2\beta$, which would only have an impact for $\tan\beta \lesssim 1$, i.e.~they can also not account for $R(K^{(*)})$~\cite{DelleRose:2019ukt,Arnan:2017lxi}.

Therefore, both charged and neutral current anomalies are challenging for the 2HDM of types I and II.
If the anomalies turn out to be true, other forms of 2HDMs with more freedom to account for the data will be needed.
In any case the exploration of the parameter space of 2HDMs, with their important interplay of 
different observables from quark and lepton flavor physics as well as high-$p_T$ measurements, will remain a cornerstone
of NP studies.

Note that in order to probe NP in $b\rightarrow c\tau\nu$, it has to be accounted for the additional complication that the measurements e.g.~of $R(D^{(*) })$ itself also depend on the specific model, see Ref.~\cite{Bernlochner:2020tfi} for details.

We follow here a model-independent way of presenting our results. 
In Sec.~\ref{sec:decayrates} we introduce the notation for differential $b\rightarrow c\tau\nu$ decay rates in the SM and scalar models, including rates for fixed $V$-polarization and fixed $\tau$-polarization, respectively. We make explicit how these decay rates are related to $b\rightarrow cl\nu$ decay rates to light leptons $l=e,\mu$.
In Sec.~\ref{sec:sumrules} we present the relations between different decay modes and derive implications for bin-wise integrated rates as well as the LFU observables $R(V)$ and $R(P)$. 
In Sec.~\ref{sec:currentdata} we give numerical results for current and hypothetical future data, after which we conclude in Sec.~\ref{sec:conclusions}.

\section{Decay Rates and Notation \label{sec:decayrates}}

\subsection{SM Decay Rates \label{sec:SMdecayrates} }

For the Standard Model (SM) expressions of $B_q\rightarrow \{V,P\}\tau \nu$ decays like 
$B\rightarrow D^{*}\tau\nu$, $B_s\rightarrow D_s^*\tau\nu$, $B_c\rightarrow J/\psi\tau\nu$ and $B\rightarrow D\tau\nu$, $B_s\rightarrow D_s\tau\nu$, $B_c\rightarrow \eta_c\tau\nu$
we employ the notation of Refs.~\cite{Bigi:2017jbd, Bigi:2017njr, Bigi:2016mdz} 
{\allowdisplaybreaks
\begin{align}
\frac{d\Gamma_{\tau,\mathrm{EXP}}^{\{V,P\}}}{dw} &= \frac{d\Gamma_{\tau,1,\mathrm{EXP}}^{\{V,P\}}}{dw} + \frac{d\Gamma_{\tau,2,\mathrm{TH}}^{\{V,P\}}}{dw}\,, \label{eq:tau-spectrum}\\
\frac{d\Gamma^{\{V,P\}}_{\tau,1,\mathrm{EXP}}}{dw} &= \left(1 - \frac{m_{\tau}^2}{q^2}\right)^2 \left(1 + \frac{m_{\tau}^2}{2q^2}\right) \frac{d\Gamma_{\mathrm{EXP}}^{\{V,P\}}}{dw}\,, \label{eq:light-leptons}  \\
\frac{d\Gamma^{V,\mathrm{TH}}_{\tau,2}}{dw} &= k P_1(w)^2\frac{m_{\tau}^2 (m_{\tau}^2 - q^2)^2 r_{V}^3 (1 + r_{V} )^2 (w^2 - 1)^{\frac{3}{2}} }{(q^2)^3 }\,, \\
 \frac{d\Gamma^{P,\mathrm{TH}}_{\tau,2}}{dw} &= k\,f_0(w)^2 \frac{m_{\tau}^2 r_P^2 (r_P^2-1)^2 \sqrt{w^2-1} (m_{\tau}^2 - m_{B_q}^2 (1+r_P^2-2 r_P w))^2  }{
	(q^2)^3
}\,, \label{eq:tau2-spectrum}
\end{align}}
where 
\begin{align}
r_{\{V,P\}} &= \frac{m_{\{V,P\}}}{m_{B_q}}\,, & 
k &= \frac{\eta_{\mathrm{EW}}^2 \vert V_{cb}\vert^2 G_F^2 m_{B_q}^5}{32\pi^3}\,, &
\eta_{\mathrm{EW}} &\simeq 1.0066\,.  
\end{align}
Here we use furthermore
\begin{align}
q^2 \equiv (p_{B_q}-p_{\{V,P\}} )^2\,, 
\end{align}
and equivalently, the dimensionless variable
\begin{align}
			w   &\equiv \frac{m_{B_{q}}^2 + m_{\{V,P\}}^2 - q^2}{2 m_{B_q} m_{\{V,P\}}} \label{eq:general-w} \\
\Leftrightarrow		q^2 &= -2 m_{B_q} m_{\{V,P\}} w  + m_{B_q}^2 + m_{\{V,P\}}^2\,.
\end{align}
The corresponding physical ranges of these are given as 
\begin{align}
m_{\tau}^2 &\leq q^2 \leq  (m_{B_q}-m_{\{V,P\}})^2\,, \\
1 &\leq w \leq \frac{m_{B_q}^2+m_{\{V,P\}}^2-m_{\tau}^2}{2 m_{B_q} m_{\{V,P\}}}\,.
\end{align}
Note that $d\Gamma/dw$ and $d\Gamma/dq^2$ are connected by the Jacobian 
\begin{align}
\left|\frac{dq^2}{dw}\right|  &= 2 m_{B_q} m_{\{V,P\}}\,, 
\end{align}
and that for different decay channels the same $q^2$ point corresponds to different $w$ points.
It is understood implicitly, that form factors of different decay modes are different.
$d\Gamma_{\tau}^{\{V,P\}}/dw$ is the decay rate spectrum with final state $\tau$ leptons, and 
$d\Gamma^{\{V,P\}}/dw$ is the one for light leptons.
We denote by the indices \lq\lq{}EXP\rq\rq{} and \lq\lq{}TH\rq\rq{} which decay rate functions are directly measurable and which are to be provided by theory. 
Of course in principle, assuming the SM, $d\Gamma^{\{V,P\}}_{\tau,2}/dq^2$ can be measured 
directly. However, for NP tests we cannot assume the SM. 
$d\Gamma_{\tau,2,\mathrm{TH}}^{\{V,P\}}$ depends on the form factors $P_1$ and $f_0$, respectively, 
which can be provided by Lattice QCD or Heavy Quark Effective Theory (HQET).  
They are related as follows to the convention of Ref.~\cite{Boyd:1997kz} (BGL), see Table~I in Ref.~\cite{Bigi:2017jbd},
\begin{align}
\mathcal{F}_2^{\mathrm{BGL}} &= \frac{1+r_{V}}{\sqrt{r_{V}}} P_1\,, &
f_0 &= f_0^{\mathrm{BGL}}/(m_{B_q}^2 - m_P^2)\,, 
\end{align}
where~\cite{Bigi:2016mdz, Boyd:1997kz} 
\begin{align}
\bra{P(p')} \bar{c} \gamma^{\mu} b\ket{\bar{B}_q(p)} &= f_+(q^2)(p+p')^{\mu} + f_-(q^2) (p-p')^{\mu}\,,  \\
\bra{V(p', \varepsilon)} \bar{c} \gamma^{\mu} b \ket{\bar{B}_q(p)} &= i g^{\mathrm{BGL}} \varepsilon^{\mu\alpha\beta\gamma} \varepsilon_{\alpha}^* p'_{\beta} p_{\gamma}\,, \\
\bra{V(p', \varepsilon)} \bar{c} \gamma^{\mu} \gamma_5 b \ket{\bar{B}_q(p)} &= f^{\mathrm{BGL}} \varepsilon^{*\mu} +
	 (\varepsilon^*\cdot p) \left[ 
        a^{\mathrm{BGL}}_+ (p+p')^{\mu} + a^{\mathrm{BGL}}_- (p-p')^{\mu}
        \right]\,, \\ 
f_0(q^2) &= f_+(q^2) + \frac{q^2}{m_{B_q}^2-m_P^2} f_-(q^2)\,,\\
m_V \mathcal{F}^{\mathrm{BGL}}_2(q^2) &= f^{\mathrm{BGL}}(q^2) + (m_{B_q}^2 - m_{V}^2) a^{\mathrm{BGL}}_+(q^2) 
	 + q^2 a^{\mathrm{BGL}}_-(q^2)\,.
\end{align}
Note that $d\Gamma^{\{V,P\}}_{\tau,1,\mathrm{EXP}}/dw$ contains only information from decays to light leptons $d\Gamma^{\{V,P\}}_{\mathrm{EXP}}/dw$, see Eq.~(\ref{eq:light-leptons}). 
The latter is given in terms of helicity amplitudes 
as~\cite{Korner:1989qb,Pham:1992fr,Fajfer:2012vx, Celis:2012dk, Bigi:2017njr}
\begin{align}
\frac{d\Gamma^V_{\mathrm{EXP}}}{dw} &= \frac{\vert V_{cb}\vert^2  G_F^2 (m_D^*)^2 q^2 \sqrt{w^2 - 1}}{48 m_B \pi^3}
		\left(H_{V,00}^2 + H_{V,--}^2 + H_{V,++}^2 \right)\,, \label{eq:light-helicities} \\
 \frac{d\Gamma^P_{\mathrm{EXP}}}{dw} &=\frac{\vert V_{cb}\vert^2  G_F^2 (m_D^*)^2 q^2 \sqrt{w^2 - 1}}{48 m_B \pi^3}
		 H_{P,0}^2   \,. 
	\end{align}
The analogous expressions for heavy final lepton states are
\begin{align}
\frac{d\Gamma^V_{\tau, \mathrm{EXP}}}{dw} &= \frac{\vert V_{cb}\vert^2  G_F^2 (m_D^*)^2 q^2 \sqrt{w^2 - 1}}{48 m_B \pi^3}
		\left(1 - \frac{m_{\tau}^2}{q^2}\right)^2 \times\nn\\
		& \left(
			\left( H_{V,00}^2 + H_{V,--}^2 + H_{V,++}^2 \right) \left( 1 + \frac{m^2_{\tau}}{2q^2}  \right) +
			\frac{3 m_{\tau}^2 }{2q^2} H^2_{V,0t}
		\right)\,, \label{eq:heavy-helicities} \\
 \frac{d\Gamma^P_{\tau, \mathrm{EXP}}}{dw} &=\frac{\vert V_{cb}\vert^2  G_F^2 (m_D^*)^2 q^2 \sqrt{w^2 - 1}}{48 m_B \pi^3}
		 \left( H_{P,0}^2 \left(1 + \frac{m^2_{\tau} }{2 q^2} \right) + \frac{3 m_{\tau}^2 }{2q^2} H_{P,0t}^2 \right)\,,
\end{align}
i.e.~$d\Gamma^{\{V,P\}}_{\tau,2}/dw$ is proportional to the additional longitudinal helicity amplitude $H_{\{V,P\},0t}^2$. 
One can measure the decay rates with a fixed $D^*$ helicity, and thereby measure each squared helicity amplitude 
in Eq.~(\ref{eq:light-helicities}) separately. We write the corresponding decay rates as
\begin{align}
\frac{d\Gamma^{V,L}_{\mathrm{EXP}}}{dw} \propto \vert H_{V,00}^2\vert^2\,,\qquad
\frac{d\Gamma^{V,T\pm}_{\mathrm{EXP}}}{dw} \propto \vert H_{V,\pm\pm}^2\vert^2\,.
\end{align} 
They fulfill by definition 
\begin{align}
\frac{d\Gamma_{\mathrm{EXP}}}{dw} &= \frac{d\Gamma^{V,L}_{\mathrm{EXP}}}{dw} + \frac{d\Gamma^{V,T+}_{\mathrm{EXP}}}{dw} + \frac{d\Gamma^{V,T-}_{\mathrm{EXP}}}{dw}\,.
\end{align}
The corresponding decay rates to $\tau$-leptons are related to those for light leptons as 
\begin{align}
\frac{d\Gamma^{V,T\pm}_{\tau,\mathrm{EXP}}}{dq^2} &= \left(1 - \frac{m_{\tau}^2}{q^2}\right)^2
				    \left(1 + \frac{m_{\tau}^2}{2 q^2}\right)  \frac{d\Gamma^{V,T\pm}_{\mathrm{EXP}}}{dq^2} \,, \\
\frac{d\Gamma^{V,L}_{\tau,\mathrm{EXP}}}{dq^2} &=  \left(1 - \frac{m_{\tau}^2}{q^2}\right)^2
			    \left(1 + \frac{m_{\tau}^2}{2 q^2}\right)  \frac{d\Gamma^{V,L}_{\mathrm{EXP}}}{dq^2}  + \frac{d\Gamma^V_{\tau,2,\mathrm{TH}}}{dq^2}\,. 
\end{align}
Similarly, for the decay rates with polarized $\tau$-leptons of helicity $\pm 1/2$ we write 
\begin{align}
\frac{d\Gamma^{\{V,P\}}_{\tau,\mathrm{EXP}}}{dw} &= \frac{d\Gamma^{\{V,P\}}_{\tau,+,\mathrm{EXP}}}{dw} + \frac{d\Gamma^{\{V,P\}}_{\tau,-,\mathrm{EXP}}}{dw}\,,
\end{align}
and where the expressions in terms of helicity amplitudes can be found in Refs.~\cite{Fajfer:2012vx,Celis:2012dk}. 
From these we can read off that 
\begin{align}
\frac{d\Gamma^{\{V,P\}}_{\tau,-,\mathrm{EXP}}}{dw} &= \left(1 - \frac{m_{\tau}^2}{q^2}\right)^2 \frac{d\Gamma^{\{V,P\}}_{\mathrm{EXP}}}{dw}\,, \\
\frac{d\Gamma^{\{V,P\}}_{\tau,+,\mathrm{EXP}}}{dw} &= \frac{m_{\tau}^2}{2 q^2} \left(1 - \frac{m_{\tau}^2}{q^2}\right)^2 \frac{d\Gamma^{\{V,P\}}_{\mathrm{EXP}}}{dw}  
+ \frac{d\Gamma^{\{V,P\}}_{\tau,2,\mathrm{TH}}}{dw} \,.
\end{align}

\subsection{Scalar Model Decay Rates \label{sec:scalardecayrates}}

For the NP part of the effective theory of a charged scalar that contributes to $b\rightarrow c\tau\nu$, we adapt the notation of Ref.~\cite{Celis:2012dk},
\begin{align}
\mathcal{L}_{\mathrm{eff}} &= -\frac{4 G_F V_{cb}}{\sqrt{2}}  
	\left(
	\bar{c} \left( C_L P_L + C_R P_R \right) b
	\right)
	\left(
	\bar{l} P_L \nu_l
	\right)\,,
\end{align}
where we implicitly use the Wilson coefficients at the $m_b$-scale.
We consider only additional scalar couplings to heavy leptons.
For sum and difference of these couplings we use the notation 
\begin{align}
\Sigma C &= C_L + C_R\,, \\
\Delta C &= C_L - C_R\,. 
\end{align}
For scalar models it is known that the only modification that enters $B_q\rightarrow V\tau\nu$ 
and $B_q\rightarrow P\tau\nu$ contribute to the longitudinal helicity amplitudes and are proportional to the  
form factors $P_1$ and $f_0$, respectively~\cite{Fajfer:2012vx, Celis:2012dk}.

The reason is that from applying the Ward identity one obtains~\cite{Fajfer:2012vx}
\begin{align}
\bra{V} \bar{c}\gamma_5 b \ket{B} &= 
\frac{1}{m_b+m_c} q^{\mu} \bra{V} \bar{c} \gamma_{\mu} \gamma_5 b \ket{B} 
= \frac{ \left(\varepsilon^*\cdot p_B\right) m_{V} }{m_b + m_c } \frac{1+r_{V}}{\sqrt{r_{V}}} P_1\,.
\label{eq:ward-1}
\end{align}
Furthermore, for $B\rightarrow D\tau \nu$ it follows~\cite{Koponen:2013tua} 
\begin{align}
\bra{P} \bar{c} \gamma_5 b \ket{B} &=  \frac{m_B^2-m_D^2}{m_b-m_c} f_0 \,. \label{eq:ward-2} 
\end{align}
Therefore, in scalar models~\cite{Fajfer:2012vx, Celis:2012dk}
\begin{align}
\frac{d\Gamma^{V,\mathrm{EXP}}_{\tau}}{dq^2} - \frac{d\Gamma^{V,\mathrm{EXP}}_{\tau,1}}{dq^2}  
 &= \frac{d\Gamma^{V,\mathrm{TH}}_{\tau,2}}{dq^2} \left| 1 - \Delta C \frac{q^2}{m_{\tau} (m_b + m_c)}\right|^2\,, \label{eq:V-scalar} \\
\frac{d\Gamma^{P,\mathrm{EXP}}_{\tau}}{dw} - \frac{d\Gamma^{P,\mathrm{EXP}}_{\tau,1}}{dw} 
&= \frac{d\Gamma^{P,\mathrm{TH}}_{\tau,2}}{dw} \left|1 + \Sigma C \frac{q^2}{m_{\tau} (m_b - m_c)}\right|^2\,, \label{eq:P-scalar}
\end{align}
where on the left hand side are only quantities that can be measured directly, whereas on the right hand side are theoretical parameters only. We have furthermore~\cite{Celis:2012dk,Murgui:2019czp,Alonso:2016oyd}
\begin{align}
\mathcal{B}(B_c\rightarrow \tau\nu) &= \mathcal{N}^{\mathrm{SM}}  
\left|1 -  r_{B_c} \Delta C  \right|^2\,, \label{eq:Bctaunu} 
\end{align}
where
\begin{align}
\mathcal{N}^{\mathrm{SM}} &\equiv \tau_{B_c} G_F^2 m_{\tau}^2 f_{B_c}^2 \vert V_{cb}\vert^2 \frac{m_{B_c}}{8\pi}\left(1 - \frac{m_{\tau}^2}{m_{B_c}^2} \right)^2\,, \label{eq:definition-norm-SM} 
\end{align}
is the SM expression for $\mathcal{B}(B_c\rightarrow \tau\nu)$ and we write
\begin{align}
r_{B_c} &\equiv \frac{m_{B_c}^2}{m_{\tau} (m_b + m_c) }\,.
\end{align}

\section{Universality Relations \label{sec:sumrules}}

\subsection{Relations for Differential Rates \label{sec:diffsumrules}}

We present now a method to differentiate between the SM and scalar models 
and compare different hadronic $b\rightarrow c\tau\nu$ decay modes in a very direct way. 
In order to do so, only theory input on the respective mode-dependent 
$d\Gamma^{\{V,P\}}_{\tau,2,\mathrm{TH}} / dq^2$ is necessary. 
Using the decay rate expressions introduced in Sec.~\ref{sec:decayrates}, that knowledge makes it possible to isolate 
$q^2$-dependent functions which do 
not depend on the concrete decay channel anymore, thereby in turn connecting different decay channels: 
\begin{align}
\forall \,\, B_q\rightarrow V\tau\nu: \quad S_{\Delta C}(q^2) &= 
	\frac{ \frac{d\Gamma_{\tau,\mathrm{EXP}}^{V}}{dq^2} - \left(1 - \frac{m_{\tau}^2}{q^2}\right)^2 \left(1 + \frac{m_{\tau}^2}{2q^2}\right) \frac{d\Gamma_{\mathrm{EXP}}^{V}}{dq^2} }{
	 \frac{d\Gamma^{V,\mathrm{TH}}_{\tau,2}}{dq^2} }\label{eq:sum-rule-10} \\
&= \frac{\frac{d\Gamma^{V,L}_{\tau,\mathrm{EXP}}}{dq^2} - \left(1 - \frac{m_{\tau}^2}{q^2}\right)^2 \left(1 + \frac{m_{\tau}^2}{2 q^2}\right)  \frac{d\Gamma^{V,L}_{\mathrm{EXP}}}{dq^2}}{ \frac{d\Gamma^{V,\mathrm{TH}}_{\tau,2}}{dq^2}} \label{eq:sum-rule-11}  \\
&= \frac{\frac{d\Gamma^V_{\tau,+,\mathrm{EXP}}}{dq^2} - \frac{m_{\tau}^2}{2 q^2} \left(1 - \frac{m_{\tau}^2}{q^2}\right)^2 \frac{d\Gamma^V_{\mathrm{EXP}}}{dq^2} }{  \frac{d\Gamma^{V,\mathrm{TH}}_{\tau,2}}{dq^2} } \,,  \label{eq:sum-rule-12}
\end{align}
and 
\begin{align}
\forall \,\, B_q\rightarrow P\tau\nu: \quad S_{\Sigma C}(q^2) &= 
	\frac{ \frac{d\Gamma^{P}_{\tau, \mathrm{EXP}}}{dq^2} - \left(1 - \frac{m_{\tau}^2}{q^2}\right)^2 \left(1 + \frac{m_{\tau}^2}{2q^2}\right)\frac{d\Gamma^{P}_{\mathrm{EXP}}}{dq^2} }{ 
	 \frac{d\Gamma^{P,\mathrm{TH}}_{\tau,2}}{dq^2} } \label{eq:sum-rule-13} \\
&= \frac{ \frac{d\Gamma^P_{\tau,+,\mathrm{EXP}}}{dq^2} - \frac{m_{\tau}^2}{2 q^2} \left(1 - \frac{m_{\tau}^2}{q^2}\right)^2 \frac{d\Gamma^P_{\mathrm{EXP}}}{dq^2} }{ \frac{d\Gamma^{P,\mathrm{TH}}_{\tau,2}}{dq^2} }\,, 	\label{eq:sum-rule-14}
\end{align} 
with the functions 
\begin{align}
S_{\Delta C}(q^2) &\equiv 1 - 2 \mathrm{Re}(\Delta C) \frac{q^2}{m_{\tau} (m_b + m_c)} + \left|\Delta C\right|^2 \left( \frac{q^2}{m_{\tau} (m_b + m_c)}\right)^2\,, \label{eq:SV-definition}\\
S_{\Sigma C}(q^2) &\equiv 1 + 2 \mathrm{Re}(\Sigma C) \frac{q^2}{m_{\tau} (m_b - m_c)} + \left|\Sigma C\right|^2 \left( \frac{q^2}{m_{\tau} (m_b - m_c)}\right)^2 \,,\label{eq:SP-definition}
\end{align}
and in the SM, trivially
\begin{align}
S_{\{\Delta C,\, \Sigma C\}}(q^2) \overset{\mathrm{SM}}{=} 1\,.
\end{align}
The slope and curvature of $S_{\Delta C}(q^2)$ and $S_{\Sigma C}(q^2)$ are directly related to scalar NP parameters.
The notation 
\lq\lq{}$\forall \,\, B_q\rightarrow V\tau\nu$\rq\rq{} and \lq\lq{}$\forall \,\, B_q\rightarrow P\tau\nu$\rq\rq{} implies that the relations hold equally for all decay channels like $B\rightarrow D^{*}\tau\nu$, $B_s\rightarrow D_s^*\tau\nu$, $B_c\rightarrow J/\psi\tau\nu$, and $B\rightarrow D\tau\nu$, $B_s\rightarrow D_s\tau\nu$, $B_c\rightarrow \eta_c\tau\nu$, respectively, with the same respective $q^2$-dependent left hand side.

\begin{figure}[t]
\begin{center}
\includegraphics[width=0.7\textwidth]{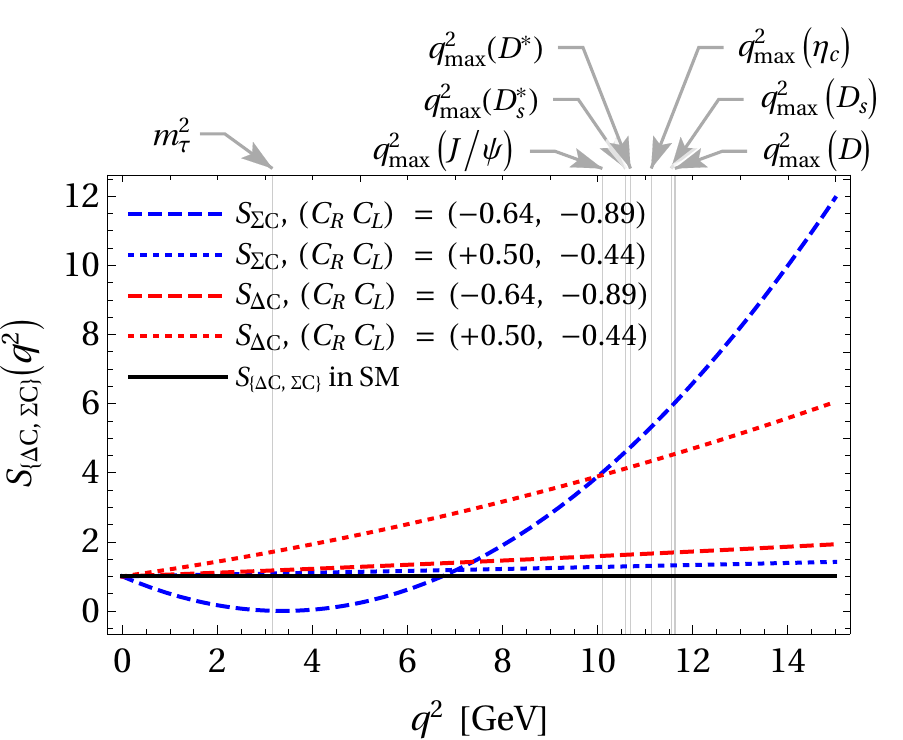} \\
\caption{The universal $q^2$-dependence $S_{\{\Delta C,\, \Sigma C\}}(q^2)$, Eqs.~(\ref{eq:SV-definition}), (\ref{eq:SP-definition}), that appears on the left-hand-side of the relations Eqs.~(\ref{eq:sum-rule-10})--(\ref{eq:sum-rule-14}) of $B_q\rightarrow \{V,P\}\tau\nu$ decays in scalar models independent of the decay mode.
The example values correspond to the minima $(C_R,C_L) = (-0.37, -0.51)$ and $(C_R,C_L) = (0.29,-0.25)$ at 1~TeV found in fits to the global $b\rightarrow c\tau\nu$ data in Table~II of Ref.~\cite{Blanke:2019qrx}, and that we 
RGE-evolve \cite{Blanke:2018yud,Gonzalez-Alonso:2017iyc} down to the $m_b$-scale. 
It is understood, that for a given decay channel the shown curve is only valid between the endpoints $m_{\tau}^2 < q^2 <  q^2_{\mathrm{max}}(\{V,P\})$, see Eq.~(\ref{eq:endpoints-2}). The region $q^2<m_{\tau}^2$ is unphysical.
From the curvature and slope of $S_{\{\Delta C,\, \Sigma C\}}(q^2)$ one can directly extract the NP parameters. 
\label{fig:sumrules}}
\end{center}
\end{figure}

Eqs.~(\ref{eq:sum-rule-10})--(\ref{eq:sum-rule-14}) are ultimately a consequence of the relation between the hadronic matrix elements Eqs.~(\ref{eq:ward-1}), (\ref{eq:ward-2}), following from the Ward identity. 
They are broken by models other than the SM and scalar models, like vector and tensor models.
Moreover, when some observables like $R(D^{(*)})$ deviate from the SM, then the simultaneous validity of the above relations  
is a hint for scalar models.
In scalar models with right-handed neutrinos, see 
Refs.~\cite{Goldberger:1999yh, Ligeti:2016npd, Iguro:2018qzf, Mandal:2020htr}, Eqs.~(\ref{eq:sum-rule-10})--(\ref{eq:sum-rule-14}) apply with modified functions $S_{\{\Delta C,\Sigma C\}}(q^2)$ that contain corresponding additional Wilson coefficients. 

In the SM-limit Eqs.~(\ref{eq:sum-rule-10}) and (\ref{eq:sum-rule-13}) trivially recover Eq.~(\ref{eq:tau-spectrum}).
Of course the division by $d\Gamma^{\{V,P\}}_{\tau,2}/dq^2$ is only possible between the endpoints of each decay channel, i.e. 
for each decay channel, Eqs.~(\ref{eq:sum-rule-10})--(\ref{eq:sum-rule-14}) are only valid for  
\begin{align}
m_{\tau}^2 &< q^2 < q^2_{\mathrm{max}}(\{V,P\})\,, \label{eq:endpoints-2}
\end{align}
which is decay-mode dependent. Note further, that the relations hold as a function of $q^2$. 
For different decay channels, a given $q^2$ point corresponds to different $w$ values, see Eq.~(\ref{eq:general-w}).
That is why we employ $d\Gamma_{\tau}^{\{V,P\}}/dq^2$ here, rather than $d\Gamma_{\tau}^{\{V,P\}}/dw$. 
In Fig.~\ref{fig:sumrules}, we show the $q^2$-dependence of Eqs.~(\ref{eq:sum-rule-10})--(\ref{eq:sum-rule-14})
for example values of $C_{L,R}$ which correspond to minima that are found in global fits to the available $b\rightarrow c\tau\nu$ decay data~\cite{Blanke:2018yud, Blanke:2019qrx}. Note that a scalar model at the scale of new physics generates 
a strongly suppressed tensor operator at the  $m_b$-scale through renormalization group equation (RGE)-running. 
We have~\cite{Gonzalez-Alonso:2017iyc, Blanke:2018yud}
\begin{align}
C_V^L(m_b) &= C_V^L(\text{1 TeV})\,, &
C_S^R(m_b) &= 1.737 \, C_S^R(\text{1 TeV})\,, \\
C_S^L(m_b) &= 1.752 \, C_S^L(\text{1 TeV})\,, &
C_T(m_b)   &= -0.004 \, C_S^L(\text{1 TeV})\,.
\end{align}
Consequently, we neglect the tensor Wilson coefficient in Fig.~\ref{fig:sumrules}. 

On top of the above relations that allow the differentiation between SM and scalar models by measuring the characteristic $q^2$-dependence, we have additional relations between the decays to $\tau$ leptons and light leptons that do not allow the differentiation between SM and scalar models, but only the one of other models from the SM and scalar models. These are 
\begin{align}
  0 &=   \frac{d\Gamma^{V,T\pm}_{\tau,\mathrm{EXP}}}{dq^2} - \left(1 - \frac{m_{\tau}^2}{q^2}\right)^2
				    \left(1 + \frac{m_{\tau}^2}{2 q^2}\right)  \frac{d\Gamma^{V,T\pm}_{\mathrm{EXP}}}{dq^2} \label{eq:sum-rule-1}\\ 
  &= \frac{d\Gamma^{\{V,P\}}_{\tau,-,\mathrm{EXP}}}{dq^2} - \left(1 - \frac{m_{\tau}^2}{q^2}\right)^2 \frac{d\Gamma^{\{V,P\}}_{\mathrm{EXP}}}{dq^2}\,.  \label{eq:sum-rule-2} 
\end{align}
Again, vector and tensor models would violate these relations. 

Eqs.~(\ref{eq:sum-rule-10})--(\ref{eq:sum-rule-14}) can also be used in order to test form factor calculations. In the ratios 
\begin{align}
\frac{d\Gamma^{\{V_1, P_1\},\mathrm{EXP}}_{\tau}/dq^2 - d\Gamma^{\{V_1,P_1\},\mathrm{EXP}}_{\tau,1}/dq^2}{
d\Gamma^{\{V_2,P_2\},\mathrm{EXP}}_{\tau}/dq^2 - d\Gamma^{\{V_2,P_2\},\mathrm{EXP}}_{\tau,1}/dq^2}
    &= \frac{d\Gamma^{\{V_1,P_1\},\mathrm{TH}}_{\tau,2}/dq^2}{d\Gamma^{\{V_2,P_2\},\mathrm{TH}}_{\tau,2}/dq^2}\,, \label{eq:lattice-check} 
\end{align}
scalar NP cancels out,~i.e. we can check the 
ratios $P_1^{V_1}(q^2)^2 / P_1^{V_2}(q^2)^2$ and $f_0^{P_1}(q^2)^2 / f_0^{P_2}(q^2)^2$ directly from data,
relying not anymore on the SM, but on the weaker assumption that at most scalar NP is present.
Of course, more general NP would invalidate this test. However, this would then also be seen in the violation of Eqs.~(\ref{eq:sum-rule-1}), (\ref{eq:sum-rule-2}). \\

Comparing to results present in the literature, the analytic relations found here are different from the numerical sum rule for the integrated observables $R(D^*)$, $R(D)$ and $R(\Lambda_c)$ in  Eqs.~(28), (29) of Ref.~\cite{Blanke:2018yud}, see also Ref.~\cite{Blanke:2019qrx}. While Eqs.~(\ref{eq:sum-rule-10})--(\ref{eq:sum-rule-14}) are model-specific, i.e.~can be used to differentiate between models, the sum rule in Refs.~\cite{Blanke:2018yud, Blanke:2019qrx} is valid for any NP model and can thus be used as a consistency check of the data. 

Eqs.~(\ref{eq:sum-rule-11}), (\ref{eq:sum-rule-12}) and (\ref{eq:sum-rule-14}) agree with the observation made in Refs.~\cite{Celis:2012dk, Celis:2016azn}, that in scalar models, the expressions including observables of one decay mode $R(D^*)(q^2)-R_L(D^*)(q^2)$ and $R(D^{(*)})(q^2) (A_{\lambda}^{D^{(*)}}(q^2)+1)$ stay SM-like, i.e.~are not suited to distinguish SM and scalar models, but only to differentiate other models. Here, $R_L(D^*)(q^2)$ is the LFU ratio of the longitudinal decay rates, and $A_{\lambda}^{D^{(*)}}(q^2)$ is the asymmetry in the $\tau$-polarization, see Refs.~\cite{Celis:2012dk,Celis:2016azn} for details.

\subsection{Relations for Integrated Rates}

\subsubsection{Bin-wise Relations \label{sec:bin-wise} }

In practice, only binned measurements of the $q^2$-dependent decay rates are performed.
The integration of the relations Eqs.~(\ref{eq:sum-rule-1}), (\ref{eq:sum-rule-2}) is straight forward. 
For Eqs.~(\ref{eq:sum-rule-10})--(\ref{eq:sum-rule-14}) there are two 
different options: (a) Integrate them in the form as written, or (b) before that multiply both sides by $d\Gamma^{\{V,P\}}_{\tau,2}/dq^2$.
Option (a) gives 
\begin{align}
& \forall \,\, B_q \rightarrow V\tau\nu: \quad \int_{\mathrm{bin}} \frac{d\Gamma^{V,\mathrm{EXP}}_{\tau}/dq^2}{d\Gamma^{V,\mathrm{TH}}_{\tau,2}/dq^2}  dq^2-
\int_{\mathrm{bin}} \frac{d\Gamma^{V,\mathrm{EXP}}_{\tau,1}/dq^2}{d\Gamma^{V,\mathrm{TH}}_{\tau,2}/dq^2} dq^2 \nn\\ 
&= 1 - 2 \mathrm{Re}(\Delta C) \int_{\mathrm{bin}}\frac{q^2}{m_{\tau} (m_b + m_c)}dq^2 + 
	\left|\Delta C\right|^2 \int_{\mathrm{bin}} \left( \frac{q^2}{m_{\tau} (m_b + m_c)}\right)^2 dq^2\,, \label{eq:integrated-sum-rules-1} \\ 
& \forall \,\, B_q\rightarrow P\tau\nu: \quad \int_{\mathrm{bin}} \frac{d\Gamma^{P,\mathrm{EXP}}_{\tau}/dq^2}{d\Gamma^{P,\mathrm{TH}}_{\tau,2}/dq^2} dq^2-
\int_{\mathrm{bin}} \frac{d\Gamma^{P,\mathrm{EXP}}_{\tau,1}/dq^2}{d\Gamma^{P,\mathrm{TH}}_{\tau,2}/dq^2} dq^2  \nn\\
&=  1 + 2 \mathrm{Re}(\Sigma C) \int_{\mathrm{bin}}\frac{q^2}{m_{\tau} (m_b - m_c)}dq^2 + \left|\Sigma C\right|^2 \int_{\mathrm{bin}} \left( \frac{q^2}{m_{\tau} (m_b - m_c)}\right)^2 dq^2\,, \label{eq:integrated-sum-rules-2}
\end{align}
and completely analogous equations for the decay rates with fixed $D^*$- and $\tau$-polarization, respectively.
We stress that it is implied that Eqs.~(\ref{eq:integrated-sum-rules-1}) and (\ref{eq:integrated-sum-rules-2}) are valid for any decay mode to vector or pseudoscalar final states, respectively, as long as on the left and the right hand side the same $q^2$-bin is considered. 
Of course it is only possible to compare bins which are kinematically accessible for each considered decay. 
Eqs.~(\ref{eq:integrated-sum-rules-1}) and (\ref{eq:integrated-sum-rules-2}) also make clear how to put the lattice form factor check Eq.~(\ref{eq:lattice-check}) into its corresponding binned version. 

In practice it is challenging to obtain the integrals in Eqs.~(\ref{eq:integrated-sum-rules-1}) and 
(\ref{eq:integrated-sum-rules-2}), 
because what actually is measured by experiment is $\int_{\mathrm{bin}} d\Gamma^{\{V,P\}}_{\tau}/dq^2\, dq^2$.
However, once the $q^2$-distribution of the above decays is measured, the evaluation of 
Eqs.~(\ref{eq:integrated-sum-rules-1}), (\ref{eq:integrated-sum-rules-2})
could be facilitated by performing the folding with the additional theory factors with the software package HAMMER~\cite{Bernlochner:2020tfi}. 

Option (b) for integrating Eqs.~(\ref{eq:sum-rule-10})--(\ref{eq:sum-rule-14}), i.e.~first multiplying by $d\Gamma^{\{V,P\}}_{\tau,2}/dq^2$, does not need this reweighting procedure. We obtain:
\begin{align}
& \forall \,\, B_q\rightarrow V\tau\nu:  \quad \int_{\mathrm{bin}} \frac{d\Gamma^{V,\mathrm{EXP}}_{\tau}}{dq^2}  dq^2-
\int_{\mathrm{bin}} \frac{d\Gamma^{V,\mathrm{EXP}}_{\tau,1}}{dq^2} dq^2 -
\int_{\mathrm{bin}} \frac{d\Gamma^{V,\mathrm{TH}}_{\tau,2}}{dq^2} dq^2\nn\\ 
&= - 2 \mathrm{Re}(\Delta C) \int_{\mathrm{bin}}\frac{q^2}{m_{\tau} (m_b + m_c)} \frac{d\Gamma^{V,\mathrm{TH}}_{\tau,2}}{dq^2}  dq^2 + 
	\left|\Delta C\right|^2 \int_{\mathrm{bin}} \left( \frac{q^2}{m_{\tau} (m_b + m_c)}\right)^2  \frac{d\Gamma^{V,\mathrm{TH}}_{\tau,2}}{dq^2}  dq^2\,, \label{eq:integrated-sum-rules-3} \\
& \forall \,\, B_q\rightarrow P\tau\nu: \quad \int_{\mathrm{bin}} \frac{d\Gamma^{P,\mathrm{EXP}}_{\tau}}{dq^2} dq^2-
\int_{\mathrm{bin}} \frac{d\Gamma^{P,\mathrm{EXP}}_{\tau,1}}{dq^2}  dq^2 - 
\int_{\mathrm{bin}} \frac{d\Gamma^{P,\mathrm{TH}}_{\tau,2}}{dq^2}  dq^2 \nn\\
&= 2 \mathrm{Re}(\Sigma C) \int_{\mathrm{bin}}\frac{q^2}{m_{\tau} (m_b - m_c)} \frac{d\Gamma^{P,\mathrm{TH}}_{\tau,2}}{dq^2}   dq^2 + \left|\Sigma C\right|^2 \int_{\mathrm{bin}} \left( \frac{q^2}{m_{\tau} (m_b - m_c)}\right)^2 \frac{d\Gamma^{P,\mathrm{TH}}_{\tau,2}}{dq^2}    dq^2\,, \label{eq:integrated-sum-rules-4}
\end{align}
and again analogous equations for the decay rates with fixed $D^*$- or $\tau$-polarization.

If the above equations are applied to multiple bins of one or several decay modes, it can be directly solved for the NP parameters.
Furthermore, it can in principle be solved for the NP parameters multiple times, generating additional relations.
In the next section we make this explicit for the case where the bin is the complete $q^2$-range.

\subsubsection{Relations for $R(V)$ and $R(P)$ \label{sec:sum-rules-R}}

We discuss now the special case of Eqs.~(\ref{eq:integrated-sum-rules-3}), (\ref{eq:integrated-sum-rules-4}) 
when the bin that we integrate over is the complete $q^2$-range.
To that end we define 
\begin{align}
R^n_{\tau, i}(\{V,P\}) &\equiv \frac{1}{\Gamma^{\{V,P\}}}\int_{m_{\tau}^2}^{(m_{B_q} - m_{\{V,P\}})^2} \left(\frac{q^2}{m_{\tau}( m_b\pm m_c)}\right)^n  \frac{d\Gamma^{\{V,P\}}_{\tau, i}}{dq^2}dq^2\,, \quad i=1,2\,,\label{eq:R-notation}\\
R_{\tau, i}(\{V,P\}) &\equiv R^0_{\tau,i}(\{V,P\}), \\
R(\{V,P\}) &\equiv R^0_{\tau}(\{V,P\}) \,,
\end{align}
with $\Gamma^{\{V,P\}}$ being the integrated decay rate for decays to light leptons, so that $R(V)$ and $R(P)$ 
are defined as usual.

Note that instead of employing the experimental measurement of $R_{\tau,1}(\{V,P\})$ from decays to light leptons, see 
Eqs.~(\ref{eq:light-leptons}), (\ref{eq:R-notation}),  we can also use the corresponding SM expression, 
because we assume the decays to light leptons to be SM-like. With
\begin{align}
R^{\mathrm{SM}}(\{V,P\}) &= R^{\mathrm{TH}}_{\tau,1}(\{V,P\}) + R^{\mathrm{TH}}_{\tau,2}(\{V,P\})\,, \label{eq:RSM}\\ 
\Delta R(\{V,P\}) &\equiv R^{\mathrm{EXP}}(\{V,P\}) - R^{\mathrm{SM}}(\{V,P\})\,,
\end{align}
and a bin over the complete $q^2$-range we have from Eqs.~(\ref{eq:integrated-sum-rules-3}), (\ref{eq:integrated-sum-rules-4}) 
\begin{align}
\forall \,\, V: \quad  \Delta R(V)
&=  -2\, R_{\tau,2}^{1,\mathrm{TH}}(V) \mathrm{Re}\left(\Delta C\right) + R_{\tau,2}^{2,\mathrm{TH}}(V) |\Delta C|^2\,, 
\label{eq:R-sum-rules-1} \\
\forall \,\, P: \quad  \Delta R(P) 
&= 2\, R_{\tau,2}^{1,\mathrm{TH}}(P) \mathrm{Re}\left(\Sigma C\right) + R_{\tau,2}^{2,\mathrm{TH}}(P) |\Sigma C|^2\,, 
\label{eq:R-sum-rules-2}
\end{align}
respectively. 
We stress again that these relations are valid for any decay mode $B_q\rightarrow V\tau\nu$ and $B_q\rightarrow P\tau\nu$, respectively. Additionally, we have from Eq.~(\ref{eq:Bctaunu})
\begin{align}
\mathcal{B}(B_c\rightarrow \tau\nu)/\mathcal{N}^{\mathrm{SM}} - 1 &= 
	-2  r_{B_c} \mathrm{Re}(\Delta C)  +
	r_{B_c}^2 \vert \Delta C\vert^2   \,. \label{eq:Bc-sum-rule}
\end{align}
Using Eqs.~(\ref{eq:R-sum-rules-1}) and (\ref{eq:R-sum-rules-2}) for multiple 
decay channels, and eliminating $\Delta C$ and $\Sigma C$, we obtain:
\begin{footnotesize}
\begin{align}
\frac{ \Delta R(V_1) }{ \Delta R(V_2) } &=  \frac{
		\left(R^{1,\mathrm{TH}}_{\tau,2}(V_3) R^{2,\mathrm{TH}}_{\tau,2}(V_1) - R^{1,\mathrm{TH}}_{\tau,2}(V_1) R^{2,\mathrm{TH}}_{\tau,2}(V_3)\right) \frac{ \Delta R(V_1)}{ \Delta R(V_3)} 
	}{  
		R^{1,\mathrm{TH}}_{\tau,2}(V_2) R^{2,\mathrm{TH}}_{\tau,2}(V_1) - R^{1,\mathrm{TH}}_{\tau,2}(V_1) R^{2,\mathrm{TH}}_{\tau,2}(V_2) +
		\left(R^{1,\mathrm{TH}}_{\tau,2}(V_3) R^{2,\mathrm{TH}}_{\tau,2}(V_2) - R^{1,\mathrm{TH}}_{\tau,2}(V_2) R^{2,\mathrm{TH}}_{\tau,2}(V_3) \right) \frac{ \Delta R(V_1)}{ \Delta R(V_3)} 
	} \label{eq:multiple-V-1} \\ 
\frac{ \Delta R(V_1) }{ \Delta R(V_2) } &= \frac{ 
	\left( r_{B_c}^2 R^{1,\mathrm{TH}}_{\tau,2}(V_1) -  r_{B_c} R^{2,\mathrm{TH}}_{\tau,2}(V_1)  \right)
	\times \frac{\Delta R(V_1)}{ \mathcal{B}(B_c\rightarrow \tau\nu)/\mathcal{N}^{\mathrm{SM}} - 1  }
	}{
		R^{1,\mathrm{TH}}_{\tau,2}(V_1) R^{2,\mathrm{TH}}_{\tau,2}(V_2) - R^{1,\mathrm{TH}}_{\tau,2}(V_2) R^{2,\mathrm{TH}}_{\tau,2}(V_1) +
		\left( r_{B_c}^2 R^{1,\mathrm{TH}}_{\tau,2}(V_2) - r_{B_c}  R^{2,\mathrm{TH}}_{\tau,2}(V_2)  \right)
		\frac{\Delta R(V_1)}{ \mathcal{B}(B_c\rightarrow \tau\nu)/\mathcal{N}^{\mathrm{SM}} - 1  }
	} \label{eq:multiple-V-2}
\end{align}
\end{footnotesize}
We can also solve directly for the NP parameters:
\begin{align}
 \vert \Delta C \vert^2 &=   \frac{
	R_{\tau,2}^{1,\mathrm{TH}}(V_1)	\Delta R(V_2)  -
	R_{\tau,2}^{1,\mathrm{TH}}(V_2) \Delta R(V_1) 
	}{ 
R_{\tau,2}^{1,\mathrm{TH}}(V_1)   R_{\tau,2}^{2,\mathrm{TH}}(V_2)  -  
R_{\tau,2}^{1,\mathrm{TH}}(V_2) R_{\tau,2}^{2,\mathrm{TH}}(V_1)
	} \\ 
&= \frac{
	\left(\mathcal{B}^{\mathrm{EXP}}(B_c\rightarrow\tau\nu)/\mathcal{N}^{\mathrm{SM}} - 1\right)  R_{\tau,2}^{1,\mathrm{TH}}(V_1) - 
	r_{B_c} \Delta R(V_1) 
	}{
	r_{B_c}^2  R_{\tau,2}^{1,\mathrm{TH}}(V_1) - r_{B_c} R_{\tau,2}^{2,\mathrm{TH}}(V_1) 
	}\,, \\
 -2\, \mathrm{Re}(\Delta C) &=  \frac{
R_{\tau,2}^{2,\mathrm{TH}}(V_2) \Delta R(V_1) 
-R_{\tau,2}^{2,\mathrm{TH}}(V_1) \Delta R(V_2) 
}{
R_{\tau,2}^{1,\mathrm{TH}}(V_1)  R_{\tau,2}^{2,\mathrm{TH}}(V_2) - R_{\tau,2}^{1,\mathrm{TH}}(V_2)  R_{\tau,2}^{2,\mathrm{TH}}(V_1)
} \\ 
&=\frac{
	r_{B_c}^2 \Delta R(V_1) 
	-\left(\mathcal{B}^{\mathrm{EXP}}(B_c\rightarrow\tau\nu)/\mathcal{N}^{\mathrm{SM}} - 1\right)  R_{\tau,2}^{2,\mathrm{TH}}(V_1) 
	}{
	r_{B_c}^2 R_{\tau,2}^{1,\mathrm{TH}}(V_1) - r_{B_c} R_{\tau,2}^{2,\mathrm{TH}}(V_1) 
	}\,. 
\end{align}
For the pseudoscalar final states it follows similarly, that 
\begin{footnotesize}
\begin{align}
\frac{\Delta R(P_1)}{\Delta R(P_2)} &= \frac{
	\left( R^{1,\mathrm{TH}}_{\tau,2}(P_3) R^{2,\mathrm{TH}}_{\tau,2}(P_1) - R^{1,\mathrm{TH}}_{\tau,2}(P_1) R^{2,\mathrm{TH}}_{\tau,2}(P_3) \right)\frac{\Delta R(P_1) }{\Delta R(P_3) }
	}{
	R^{1,\mathrm{TH}}_{\tau,2}(P_2) R^{2,\mathrm{TH}}_{\tau,2}(P_1) - R^{1,\mathrm{TH}}_{\tau,2}(P_1) R^{2,\mathrm{TH}}_{\tau,2}(P_2) +
	\left( R^{1,\mathrm{TH}}_{\tau,2}(P_3) R^{2,\mathrm{TH}}_{\tau,2}(P_2) - R^{1,\mathrm{TH}}_{\tau,2}(P_2) R^{2\,\mathrm{TH}}_{\tau,2}(P_3)  \right)
	 \frac{\Delta R(P_1) }{\Delta R(P_3) }
	}\,, \label{eq:multiple-P} 
\end{align}
\end{footnotesize}
\begin{align}
 \vert \Sigma C \vert^2 &= \frac{
	R_{\tau,2}^{1,\mathrm{TH}}(P_1)	\Delta R(P_2)  -
	R_{\tau,2}^{1,\mathrm{TH}}(P_2) \Delta R(P_1)  
	}{ 
R_{\tau,2}^{1,\mathrm{TH}}(P_1)   R_{\tau,2}^{2,\mathrm{TH}}(P_2)  -  
R_{\tau,2}^{1,\mathrm{TH}}(P_2) R_{\tau,2}^{2,\mathrm{TH}}(P_1)
	}\,,   \\\nn\\
 2\, \mathrm{Re}(\Sigma C) &=  \frac{
R_{\tau,2}^{2,\mathrm{TH}}(P_2) \Delta R(P_1) 
-R_{\tau,2}^{2,\mathrm{TH}}(P_1) \Delta R(P_2) 
}{
R_{\tau,2}^{1,\mathrm{TH}}(P_1)  R_{\tau,2}^{2,\mathrm{TH}}(P_2) - R_{\tau,2}^{1,\mathrm{TH}}(P_2)  R_{\tau,2}^{2,\mathrm{TH}}(P_1)
}  
\end{align}
Analogous relations can be obtained for fixed $\tau$ or $V$ polarization.

\subsubsection{Approximate Relations \label{sec:approx}}

In the limit of a small NP contribution, i.e.~in case that 
\begin{align}
\frac{\vert R_{\tau,2}^{2,\mathrm{TH}}(V)\vert}{2 \vert R_{\tau,2}^{1,\mathrm{TH}}(V)\vert} \frac{ \vert \Delta C\vert^2 }{ \vert \mathrm{Re}(\Delta C)\vert}  \ll 1\,, \label{eq:approx-condition-1}\\
\frac{1}{2} r_{B_c} \frac{ \vert \Delta C\vert^2 }{\vert \mathrm{Re}(\Delta C)\vert} \sim 2 \frac{ \vert \Delta C\vert^2 }{\vert \mathrm{Re}(\Delta C)\vert} \ll 1\,, \label{eq:approx-condition-2}
\end{align}
we find approximate relations that are simpler than the ones derived in Sec.~\ref{sec:sum-rules-R}. 
From Eqs.~(\ref{eq:R-sum-rules-1}), (\ref{eq:Bc-sum-rule}) we have in this case 
\begin{align}
\frac{\Delta R(V_1)}{\Delta R(V_2)} &= \frac{R_{\tau,2}^{1,\mathrm{TH}}(V_1)}{ R_{\tau,2}^{1,\mathrm{TH}}(V_2)}\,, \label{eq:approx-multiple-V-1} 
\end{align}
and 
\begin{align}
 -2\, \mathrm{Re}\left(\Delta C\right) &= \frac{\Delta R(V)}{ R_{\tau,2}^{1,\mathrm{TH}}(V)}  
= \frac{1}{r_{B_c}} \left(\frac{\mathcal{B}(B_c\rightarrow \tau\nu)}{\mathcal{N}^{\mathrm{SM}}} - 1\right)\,.  \label{eq:sum-rule-approx-integrated}
\end{align}
When $\vert \Delta C\vert^2$ is not known, a check of Eqs.~(\ref{eq:approx-condition-1}) and 
(\ref{eq:approx-condition-2}) is not available. 
However, the conditions Eqs.~(\ref{eq:approx-condition-1}), (\ref{eq:approx-condition-2}) also imply the weaker inequalities 
\begin{align}
\frac{\vert R_{\tau,2}^{2,\mathrm{TH}}(V)\vert}{2 \vert R_{\tau,2}^{1,\mathrm{TH}}(V)\vert} 
\vert \mathrm{Re}(\Delta C)\vert  \ll 1\,, \label{eq:consistency-check-1}\\
\frac{1}{2} r_{B_c} \vert \mathrm{Re}(\Delta C)\vert \sim 2 \vert \mathrm{Re}(\Delta C)\vert \ll 1\,, 
\label{eq:consistency-check-2}
\end{align}
which can be used for a consistency check after the extraction of $\mathrm{Re}(\Delta C)$ through
Eq.~(\ref{eq:sum-rule-approx-integrated}).

Analogously, for semileptonic decays to pseudoscalars we have the approximate relations
\begin{align}
\frac{\Delta R(P_1)}{\Delta R(P_2)} &= \frac{ R_{\tau,2}^{1,\mathrm{TH}}(P_1)  }{ R_{\tau,2}^{1,\mathrm{TH}}(P_2)}\,, \label{eq:approx-multiple-P} \\
 2\, \mathrm{Re}\left(\Sigma C\right)  &= \frac{\Delta R(P)}{ R_{\tau,2}^{1,\mathrm{TH}}(P)} \,,  \label{eq:sum-rule-approx-integrated-P}
\end{align}
which are valid if the relation
\begin{align}
\frac{\vert R_{\tau,2}^{2,\mathrm{TH}}(P)\vert}{2 \vert R_{\tau,2}^{1,\mathrm{TH}}(P)\vert } \frac{ \vert \Sigma C\vert^2 }{ \vert \mathrm{Re}(\Sigma C)\vert} \leq 
\frac{\vert R_{\tau,2}^{2,\mathrm{TH}}(P)\vert}{ 2\vert R_{\tau,2}^{1,\mathrm{TH}}(P)\vert} \vert\mathrm{Re}(\Sigma C) \vert
	\ll 1
\end{align}
is fulfilled.

\section{Application to Data \label{sec:currentdata}}

\subsection{Current Data}

\subsubsection{Relations between $\mathcal{B}(B_c\rightarrow \tau \nu)$, $R(D^*)$ and $R(J/\psi)$ for small scalar NP \label{sec:application-approx}} 

We apply the relations of Sec.~\ref{sec:sumrules} to the current measurements of charged current LFU observables
that we list in Sec.~\ref{sec:intro}. 
With current data we can test the approximate relation Eq.~(\ref{eq:sum-rule-approx-integrated}) for $V=D^*$ and $V=J/\psi$. 
Note that no direct measurement of $R^{\mathrm{EXP}}_{\tau,1}(D^*)$ is available, so that we use its SM value, 
see Eq.~(\ref{eq:RSM}). 
We use the fit results for $B\rightarrow D^*\tau\nu$ from Ref.~\cite{Gambino:2019sif}, including $R(D^*)^{\mathrm{SM}}$ as given in Eq.~\ref{eq:SM-value}, which employs recent data on decays to light leptons~\cite{Amhis:2019ckw, Abdesselam:2017kjf, Waheed:2018djm}, as well as HQET input for $P_1$, 
see Ref.~\cite{Gambino:2019sif} for details.
For the needed integrals we obtain 
\begin{align}
R_{\tau,2}^{1,\mathrm{TH}}(D^*) &= 0.018^{+0.005}_{-0.004}\,, \label{eq:rdstar1tau2-result}\\
R_{\tau,2}^{2,\mathrm{TH}}(D^*) &= 0.013 \pm 0.003 \,. \label{eq:rdstar2tau2-result} 
\end{align}
For $B_c\rightarrow J/\psi\tau\nu$ we use Eqs.~(\ref{eq:current-data-3}), (\ref{eq:RJpsi-SM}) and the fit results provided in Ref.~\cite{Cohen:2019zev}.
We obtain for the needed integrals 
\begin{align}
R^{1,\mathrm{TH}}_{\tau,2}(J/\psi) &= 0.017 \pm 0.005\,,\label{eq:rJpsi1tau2-result} \\
R^{2,\mathrm{TH}}_{\tau,2}(J/\psi) &= 0.012^{+0.004}_{-0.003}\,. \label{eq:rJpsi2tau2-result} 
\end{align}
Therein, we also take into account the correlations between the $z$-expansion coefficients of the form factors of $B_c\rightarrow J/\psi\tau\nu$ provided in Ref.~\cite{Cohen:2019zev}, however, we do not take into account further correlations like with the form factor coefficients of $B\rightarrow D^*\tau\nu$.
Note that our input from Refs.~\cite{Gambino:2019sif, Cohen:2019zev} takes into account statistical and systematic errors, and so do consequently also our numerical results. 
We use furthermore~\cite{Colquhoun:2015oha} 
\begin{align}
f_{B_c} = (0.434\pm 0.015)\, \mathrm{GeV}\,.
\end{align}
The approximate relation Eq.~(\ref{eq:sum-rule-approx-integrated}) implies
\begin{align}
\mathcal{B}(B_c\rightarrow \tau\nu) &=  \mathcal{N}^{\mathrm{SM}} \left(1 + r_{B_c} \frac{\Delta R(D^*)}{R_{\tau,2}^{1,\mathrm{TH}}(D^*)} \right)\,, \label{eq:prediction-Bc} \\ 
R(J/\psi) &= R(J/\psi)^{\mathrm{SM}} + \Delta R(D^*) \frac{R_{\tau,2}^{1,\mathrm{TH}}(J/\psi)}{R_{\tau,2}^{1,\mathrm{TH}}(D^*)}\,, \label{eq:approx-RJpsi}  
\end{align}
see Eq.~(\ref{eq:definition-norm-SM}) for the definition of $\mathcal{N}^{\mathrm{SM}}$.
Before we evaluate these expressions numerically, we perform the consistency check 
Eq.~(\ref{eq:consistency-check-2}) required for actually applying the used approximation from Sec.~\ref{sec:approx}.
We obtain 
\begin{align}
\mathrm{Re}(\Delta C) &= -\frac{1}{2}\frac{\Delta R(D^*)}{R_{\tau,2}^{1,\mathrm{TH}}(D^*)} 
	= -1.1^{+0.5}_{-0.7}\,. \label{eq:redeltac} 
\end{align}
Note that $\mathrm{Re}(\Delta C)$ in Eq.~(\ref{eq:redeltac}) is large and actually violates the consistency check, therefore invalidating Eqs.~(\ref{eq:prediction-Bc})--(\ref{eq:redeltac}). 
In the next section we therefore consider a relation that does not rely on the approximation of small Wilson coefficients.

\subsubsection{Relation between $\mathcal{B}(B_c\rightarrow \tau \nu)$, $R(D^*)$ and $R(J/\psi)$ for arbitrary scalar NP \label{sec:application-exact}} 

As described in Sec.~\ref{sec:application-approx}, with current data the approximate relation between $\mathcal{B}(B_c\rightarrow \tau \nu) $ and $R(D^*)$ is not applicable, because $\vert \Delta C\vert^2$ turns out to be too large.
Consequently, instead of the approximate relations from Sec.~\ref{sec:approx}, we need to use the exact relations from Sec.~\ref{sec:sum-rules-R}. We have   
\begin{footnotesize}
\begin{align}
& R(J/\psi) = R(J/\psi)^{\mathrm{SM}} + \nn\\ & \frac{
	\left(\frac{\mathcal{B}(B_c\rightarrow \tau\nu)}{\mathcal{N}_{\mathrm{SM}}} - 1 \right) (R_{\tau,2}^{1,\mathrm{TH}}(J/\psi) R_{\tau,2}^{2,\mathrm{TH}}(D^*) - R_{\tau,2}^{1,\mathrm{TH}}(D^*) R_{\tau,2}^{2,\mathrm{TH}}(J/\psi)) + 
	r_{B_c} \Delta R(D^*) (R_{\tau,2}^{2,\mathrm{TH}}(J/\psi) - r_{B_c} R_{\tau,2}^{1,\mathrm{TH}}(J/\psi) )
	}{
	r_{B_c} (R_{\tau,2}^{2,\mathrm{TH}}(D^*) - r_{B_c} R_{\tau,2}^{1,\mathrm{TH}}(D^*))
	}\,. \label{eq:RJpsi-prediction} 
\end{align}
\end{footnotesize}
As input for $B\rightarrow D^*\tau\nu$ and $B_c\rightarrow J/\psi\tau\nu$ we employ again the fit results 
from Refs.~\cite{Gambino:2019sif, Cohen:2019zev}.
Furthermore, we vary $\mathcal{B}(B_c\rightarrow \tau\nu)$ in the conservative region $0\leq \mathcal{B}(B_c\rightarrow \tau\nu)\leq 0.6$~\cite{Blanke:2018yud,Blanke:2019qrx}, see also Refs.~\cite{Murgui:2019czp, Alonso:2016oyd, Celis:2016azn, Beneke:1996xe, Akeroyd:2017mhr, Acciarri:1996bv}.
We use the fit results Eqs.~(\ref{eq:rdstar1tau2-result})--(\ref{eq:rJpsi2tau2-result}) in Eq.~(\ref{eq:RJpsi-prediction}) and for simplicity use Gaussian error propagation without correlations to calculate the error of $R(J/\psi)$. We obtain thereby the scalar model prediction
\begin{align}
R(J/\psi) &= 0.29\pm 0.04\,, \label{eq:num-prediction-Jpsi} 
\end{align}
which has a $1.7\sigma$ tension with the current measurement Eq.~(\ref{eq:current-data-3}).

\subsection{Future Data Scenario \label{sec:future-data}}

In order to further explore the implications of Eq.~(\ref{eq:RJpsi-prediction}), we consider a 
hypothetical future data set given in Table~\ref{tab:future-data}, and motivated from prospects at Belle~II and LHCb. 
At 50~ab$^{-1}$ Belle~II expects a relative error on $R(D^*)$ of  $(\pm 1.0\pm 2.0)\%$, see Table~50 in Ref.~\cite{Kou:2018nap}.
At 50~fb$^{-1}$ LHCb expects an absolute precision, combining statistical and systematical errors, for $R(D^*)$ of $\sim 0.006$ and for $R(J/\psi)$ of $\sim 0.05$, see Fig.~55 in Ref.~\cite{Cerri:2018ypt}.
With the input of $R(D^*)$ from Table~\ref{tab:future-data}, we find the prediction Eq.~(\ref{eq:num-prediction-Jpsi}) almost unchanged,
\begin{align}
R(J/\psi) &= 0.29\pm 0.03\,.
\end{align}
This highlights the importance of a future improvement of the theory uncertainty of the scalar form factors. However, the deviation of $R(J/\psi)^{\mathrm{EXP}}$ as given in Table~\ref{tab:future-data} would amount in this scenario to an exclusion of scalar models by $7.2\sigma$. 

Note that with future data of course many more opportunities arise to apply the methods presented above, when the spectrum of $b\rightarrow c\tau\nu$ decays is measured. This will further enhance the possible significances for the exclusion of models, as well as the ability to detect NP.

\begin{table}[t]
\begin{center}
\begin{tabular}{c|c}
\hline \hline
Observable & Hypothetical Future Data \\\hline

$R(J/\psi)^{\mathrm{EXP}}$ &  $0.71 \pm 0.05$ \\ 

$R(D^*)^{\mathrm{EXP}}$    & $0.295\pm 0.006$ \\\hline\hline                      
\end{tabular}
\caption{Future scenario for hypothetical experimental data, with combined statistical and systematic errors.
\label{tab:future-data}}
\end{center}
\end{table}

\section{Conclusions \label{sec:conclusions}}

We find relations between differential decay rates of different $b\rightarrow c\tau\nu$ decay modes in scalar models.
The relations are given in Eqs.~(\ref{eq:sum-rule-10})--(\ref{eq:sum-rule-14}) and  
show a universal $q^2$-dependence for all decay modes to vector and  pseudoscalar final states, respectively.
They follow ultimately from the Ward identity for scalar hadronic matrix elements. 
Models different from scalar models break the relations. Requiring only theoretical knowledge on the scalar 
$B\rightarrow \{V,P\}$ form factor, and otherwise only experimental measurements of various decay rates and their 
phase space weighted form, 
it is possible to disentangle Standard Model (SM) and scalar models by determining the characteristic decay mode independent  
function $S_{\{\Delta C,\, \Sigma C\}}(q^2)$, see Eqs.~(\ref{eq:SV-definition}), (\ref{eq:SP-definition}), that we show in Fig.~\ref{fig:sumrules}. 

From the slope and curvature of $S_{\{\Delta C,\, \Sigma C\}}(q^2)$ one can directly extract new physics parameters. The SM-limit is given by $S_{\Delta C}(q^2)=S_{\Sigma C}(q^2)=1$. Furthermore, the cancellation of scalar new physics in the ratio Eq.~(\ref{eq:lattice-check}) allows for a check of lattice results for scalar form factors. The check does not rely on the SM, but on the weaker assumption that at most scalar new physics is present. Signatures of other new physics models would also be seen in the violation of other relations, 
like Eqs.~(\ref{eq:sum-rule-1}), (\ref{eq:sum-rule-2}). 

We make explicit the implications for corresponding bin-wise integrated rates as well as for ratios 
of $\Delta R(\{V,P\})\equiv R^{\mathrm{EXP}}(\{V,P\})-R(\{V,P\})^{\mathrm{SM}}$ for different decay channels, see
Eqs.~(\ref{eq:multiple-V-1}), (\ref{eq:multiple-V-2}), (\ref{eq:multiple-P}). 
For small new physics Wilson coefficients, i.e.~in case their second order contribution is negligible, we obtain the simpler
approximate relations Eqs.~(\ref{eq:approx-multiple-V-1}), (\ref{eq:sum-rule-approx-integrated}), (\ref{eq:approx-multiple-P}).
We note that a generalization of these results to $b\rightarrow u\tau\nu$ decays seems straight forward. 

Note that in case the anomalies turn out to be a statistical fluctuation, the 2HDM type~II would again be a very important and viable candidate for further studies. In that case, and disregarding the flipped sign solution, Higgs data shows that we are close to the alignment limit $\cos(\beta-\alpha)=0$, see the constraints on the parameter space of $\tan\beta$ vs. $\cos(\beta-\alpha)$ from ATLAS and CMS Run I+II in Fig.~11 of Ref.~\cite{Kraml:2019sis}. However, without imposing a symmetry it would actually be unnatural if the alignment limit was fulfilled exactly, which raises the interest in the parameter space with $1\%\lesssim \cos(\beta-\alpha) \lesssim 10\%$ and the corresponding more stringent bounds in that region, roughly overall about $\tan\beta \lesssim 15$. 

Future experimental results will show if the charged current anomalies are indeed true. 
With future theoretical results on the scalar form factors from lattice QCD as well as experimental measurements of the $q^2$-dependence of $b\rightarrow c\tau\nu$ decays, using the above methodology we will then be able to improve the probes for new physics in a very direct and clear way. 

\begin{acknowledgments}
We thank Paolo Gambino, Martin Jung, Henry Lamm and Richard Lebed for discussions.
The work of A.S.~is supported in part by the US DOE Contract No. DE-SC 0012704.
S.S.~is supported by a DFG Forschungs\-stipendium under contract no. SCHA 2125/1-1.
\end{acknowledgments}

\bibliography{draft.bib}
\bibliographystyle{apsrev4-1}

\end{document}